# Instrumented Collective Learning Situations (ICLS): the gap between theoretical research and observed practices

Michel Christine, Garrot Elise, George Sébastien
Laboratoire LIESP, INSA-Lyon
21, avenue Jean Capelle – F-69621 Villeurbanne Cedex, France
*Christine.Michel@insa-lyon.fr, Elise.Garrot@insa-lyon.fr, Sebastien.George@insa-lyon.fr*

**Abstract**: According to socio-constructivism approach, collective situations are promoted to favor learning in classroom, at a distance or in a blended educational context. So, many Information and Communication Technologies (ICT) are provided to teachers but there are no clear studies about the way they are used and perceived. Our research is based on the hypothesis that practices of educational actors (instructional designers and tutors) are far away from theoretical results of research in education technologies. In this paper, we consider a precise kind of situation: Instrumented Collective Learning Situations (ICLS). By a survey on 13 fields in higher education in France, Switzerland and Canada, we present how ICLS are designed and how teachers used them. Conclusions give an indication on the gap between the way information technologies are prescribed and the way they are actually used and perceived by teachers.

## Introduction

Numerous educational platforms exist to support distance or blended education. Most of them clearly announce the desire to favour collective learning. In this sense, these platforms provide several communication tools like discussion forum, chat or wiki. Nevertheless, to only provide communication tools is not always sufficient to create interactions between learners and to favour the construction of collective knowledge. The setting-up of a collective learning activity is the central point to support a socio-constructivist approach (Doise & Mugny 1984). To simply offer technical possibilities for communication would appear to be insufficient to ensure collective learning. Although the learners are offered a possibility to communicate, there is no reason for them to do so if there is no common interest bringing them together.

First works on educational platforms have focused on resources to give to learners. Since, research works have moved towards the support of modelling and designing collaborative learning activities. As a result, the definition of an Educational Modelling Language EML (Koper 2000) leads to the standard IMS-LD (2003), in which more and more platforms are compliant with. However, we think that currently an important gap remains between researches and realities in the field of education that used computing technologies. In practice, teachers who want to set up collaborative activities between students often use a pragmatic approach. Even if some specific universities are specialized in distance education, for instance the Open University and the Tele-University of Quebec, most universities try to add distance programs or instrumented activities to face to face teaching. Teachers implied in this kind of experience are often few trained and they adopt a very pragmatic attitude, far away from research works.

The work presented in this article, done in the context of the ACTORS TICE project, concerns the observation and analysis of Instrumented Collective Learning Situations (ICLS) in higher education. We define an ICLS as an educational situation with a learning goal (of knowledge or competencies), involving identified actors during a specified period and with an assessment of learners. This situation takes the shape of a scripted learning unit in which the expected production is bound to a collective activity instrumented by computer artefacts. Main goals of ICLS observation and analysis are : 1) to observe the reality in order to evaluate the gap between research and practices, 2) to provide guidelines basing on a bottom-up approach (defining best practices). In this paper we present results about a survey where instructional designer and tutors are asked about there ICLS e-learning experiment.

## Context of research

Educational institutions tend to provide diversified learning activities for students, like project-based learning, business games or case studies. This kind of activities has already proved to be useful. According to the socio-constructivist approach, interactions between learners play a dynamic role in individual learning

(Dillenbourg 1999). Nowadays, collective learning activities are more and more instrumented, in particular in order to be carried out by students at a distance, with the monitoring of tutors. Students and tutors need at least communication tools like forum or chat. Instrumentation of these activities allows passing from presence to distance and gives new possibilities like individualization and flexibility in terms of space and time. Institutions can so get a new public, like persons with disabilities and working people. Furthermore, teachers use more and more ICT due to instructional reasons or extern pressures (academic, economic, social).

Realizing the prominent part played by ICT, and their use in collective learning activities, we decide to study some ICLS in higher education context. The term 'collective' in ICLS includes both concepts of 'collaborative' and 'cooperative' (George & Leroux 2001). An ICLS can have various levels of granularity, it can thus be a session, a sequence or a learning unit, since it is bounded by the characteristics expressed in the definition we give in the introduction (duration, learning objective, individual or collective production). A production is any observable result of an activity, for example: an individual or collective synthesis, texts recorded from communication tools (e.g. forum, e-mail, and chat), connection tracks… ICLS description criterions are very precise. It allows qualifying, analyzing and comparing specific fields of experiment. We distinguish ICLS from CSCL (Computer Support for Collaborative Learning) (Bannon 1989). CSCL refers to a research field related to learning through collaborative activity and to the exploration of how such learning might be augmented through technology (ISLS 2006). So CSCL works often aim to design collaborative environments which exercise an active control over the collaborative interactions (Jermann 2001), that is not necessarily a condition in what we name ICLS.

Our research considers two different roles for teachers: instructional designers of situations (scenarios, activities, contents), and tutors who monitor learners to help them to build knowledge and competencies and to assess them. Indeed, the use of ICT in education involved a new definition of teacher roles, even if up to now these are rather badly defined and vary from one educational institution to the other (Garrot 2006). For the course to be effective, Mc Pherson & Nunes (2004) insist on the importance of both tutors and instructional designers' roles: "*the focus is frequently placed on design and developing information and communication technology (ICT) based environments and insufficient attention is given to the delivery process.*" One of the most important points of our study is to determine in which proportion ICLS really exists in higher education and how ICT are used. It is not because the activity is instrumented that actors are effectively going to use tools. For example, according to Casey & Greller (2005), tutors try to adapt given scripted courses to their students without help, building their own experiences. That is why we are interested in studying the effective use of every tools offered to support the design and the progress of CLS.

## Research issues

Our research is based on the hypothesis that educational actors' practices (designers and tutors) are far away from research in education technologies. Concerning instructional designers, many current researches are centred on the characterization and the standardization of learning activities (like IMS LD) in order to develop educational software to assist the design of scenarios. These tools aim at helping designers in the setting-up and adaptation of learning scenarios (for example to create an activity from a generic model) or at improving the scenario by visualizing uses tracks. But in practice, designers do not strictly formalize scenarios and most of them do not use developed educational software. Furthermore, ICT companies provide learning platforms which offer more and more possibilities for tutors and learners with several tools: communication tools (chat, email, forum, videoconferencing, audioconferencing), sharing/production tools (application sharing, whiteboard, shared text editor, wiki, blog, portfolio) and collaborative work management tools (planning, documents management). But learners and tutors, in their day-to-day practice, do not use all technologies that are offered to them.

In this study, we make the hypothesis that there is a considerable difference between what is prescribed and what is used. We distinguish designers' practices and those of tutors and learners because there are two kinds of prescription. On the one hand, we can ask if designers have adequate platforms to design collective learning activities according to instructional specialists' prescriptions. On the other hand, we would like to know if tutors and learners have adapted tools to carry out CLS and if they use them in the way prescribed by the designer or not (Rabardel 1995). This study tends to evaluate the gap between research and software, which provide actors with many guidelines, and the reality of actors' practices. According to our hypothesis, we make an investigation in several fields in higher education which design and use ICLS. This investigation is supported by a questionnaire that we develop in the next part.

## Methodology

A survey was elaborated starting from a list of 69 criterion defined in the ACTORS TICE project by a group of 8 researchers (Bourriquen *et al.* 2006). 207 questions have been defined to identify practices of three actors' types: instructional designers, tutors and learners. When it is pertinent, distinction is always made between what have been prescribed (in terms of tools and scenarios for example) and what have been effectively done. We collect in particular variation of practices concerning communication tools, sharing/production tools and collective work management tools. The questionnaire used in the survey is composed with a part of common queries and a part of specific questions depending on actors' types. Common questions relate to global context description and learning tools (context of execution, teaching objectives, number of instructional designers, tutors and learners participating, tools provided and used, kind of collaboration, resources available, exchanges control and learning assessment). Addressing these questions to several actors makes it possible to reinforce answers validity and identify possible perceptions variations. That also allows identifying variations of practices as we will specify further. Specific questions, reserved to each actors, relates to their perception of tools that they have to use. They are asked for example about (1) tools which are used in a wrong or in another way that the one initially defined, (2) helps and constraints felt by using the learning tools, (3) waiting and evolution proposals. In the ACTORS TICE project, various experimentation fields, located in France, Switzerland and Canada (Québec), have been studied: distance campus FORSE, Did@ctic and VCIEL; TECFA University (UTICEF/MALTT), University of Lyon 1 (E-Miage and FISAD), INSA of Lyon (Business games), Savoy University (English formation), Central school of Lyon (EAT), Teluq-UQAM (IAO formation). They dispense distance, presence or blended courses. Disciplines concerned are variable (language sciences, education sciences, data processing, management…). There did not exist as many ICLS as initially supposed on the survey because most of teaching education models are based on individual rather than collective tasks.

Results presented here concern 18 ICLS observations composed of instructional designers and tutors answers. Answers have been analysed as follow. Closed questions were processed according to simple statistical analysis (counting, average, frequencies) to have a global view of the context and technical learning tools (provided and used) in ICLS and cross together to see how they are used. Open questions have not been considered at this step of analyse.

## Results

### Activities types

Among the collective activities types offered to learners, we find information retrieval, debate, collective reading/writing, working in project, case studies and problem solving (see table 1). We do not observe prominent trend, all activities types are offered in a distributed way.

| % | Information retrieval | Debate | Reading/Writing | Project | Case studies | Problem solving |
|---|---|---|---|---|---|---|
| No or rather no | 50 | 38,9 | 33,3 | 44,4 | 44,4 | 50 |
| Yes or rather yes | 50 | 61,1 | 66,7 | 55,6 | 55,6 | 50 |

**Table 1**: Type of activities offered to learners (in % of observed ICLS).

### Cooperative versus collaborative

The proportion of collective activities prescribed by the instructional designer is high (64,38%, the balance being individual activities which participate to the collective work). It is massively follows because we find 60,71% of collective activities in practice. The answers mention that collective work is rather cooperative (50,00% in prescription and 59,64% in practice) than collaborative. These results are to be considered with caution because the notion of cooperative-collaborative is not clearly spread or not formalized among the actors questioned in our survey (even if a definition was given).

### Synchronous and asynchronous activities

Table 2 presents the percentage of synchronous and asynchronous activities carried out by tutors and learners, as they were prescribed by the instructional designer or not. In a synthetic way, we can say that both types of activities are often prescribed, although synchronous are more prescribed than asynchronous ones for learners (87,5% vs. 68,75%) and for tutors (100% vs. 64,29%). But we can see that, when they are prescribed, asynchronous activities are used while synchronous activities are rather often little used (approximately one third of prescribed synchronous activities is little used).

| Learners' activities | Prescribed and not used | Prescribed and a little used | Prescribed and often used | Not prescribed and not used | Not prescribed and a little used | Not prescribed and often used |
|---|---|---|---|---|---|---|
| Synchronous activities | 0 | 31,25 | 56,25 | 0 | 12,50 | 0 |
| Asynchronous activities | 0 | 6,25 | 62,50 | 31,25 | 0 | 0 |
| **Tutors' activities** | **Prescribed and not used** | **Prescribed and a little used** | **Prescribed and often used** | **Not prescribed and not used** | **Not prescribed and a little used** | **Not prescribed and often used** |
| Synchronous activities | 0 | 33,33 | 66,67 | 0 | 0 | 0 |
| Asynchronous activities | 0 | 0 | 64,29 | 35,71 | 0 | 0 |

**Table 2**: Synchronous and asynchronous activities for learners and tutors (in % of observed ICLS)

**Tools provided and used by tutors and learners**

In the following 3 tables, we present percentage of tools prescribed and used by learner and tutor. In the learner case the prescription is made by tutor. In the tutor case the prescription is made by instructional designer.

Table 3 shows that the massively provided communication tools are e-mail (68,75% for learners and 76,47% for tutors) and forum (68,75% for learners and 64,7% for tutors) and chat (56,25% for learners and 58,82% for tutors) (cf. table 3). When these tools are provided, they are mostly used. We said that synchronous activities are very often prescribed but they are electronically supported only by chat, which is rather little used. This can explain why provided synchronous activities are rather little used. Learners use instant messaging tools which are not provided on the learning environment (like MSN Messenger). For asynchronous activities, learners use more forum than e-mail while tutors prefer e-mail.

Phone solution is rarely offered to learners (12,50%) and even if it is often offered to tutors (47,05%) they only use it a little. Few platforms provide audio or visio conferencing tools and when they are provided they are little used. Face to face communication is possible in 40% of ICLS for learners and 57,14% for tutors.

| Learners' uses | Provided and not used | Provided and a little used | Provided and often used | Not provided and not used | Not provided and a little used | Not provided and often used |
|---|---|---|---|---|---|---|
| e-mail | 0 | 37,50 | 31,25 | 31,25 | 0 | 0 |
| forum | 0 | 18,75 | 50,00 | 31,25 | 0 | 0 |
| Chat | 0 | 31,25 | 25,00 | 31,25 | 12,50 | 0 |
| Visio conferencing | 0 | 12,50 | 0 | 81,25 | 0 | 6,25 |
| Audio conferencing | 0 | 0 | 0 | 100 | 0 | 0 |
| Phone | 0 | 12,50 | 0 | 81,25 | 6,25 | 0 |
| Mail | 6,25 | 6,25 | 0 | 81,25 | 6,25 | 0 |
| Face-to-face | 0 | 6,67 | 33,33 | 60,00 | 0 | 0 |
| **Tutors' uses** | **Provided and not used** | **Provided and a little used** | **Provided and often used** | **Not provided and not used** | **Not provided and a little used** | **Not provided and often used** |
| e-mail | 0 | 35,29 | 41,18 | 23,53 | 0 | 0 |
| forum | 5,88 | 23,53 | 35,29 | 35,29 | 0 | 0 |
| Chat | 5,88 | 29,41 | 23,53 | 29,41 | 11,76 | 0 |
| Visio conferencing | 0 | 17,65 | 0 | 76,47 | 5,88 | 0 |
| Audio conferencing | 0 | 11,76 | 0 | 88,24 | 0 | 0 |
| Phone | 0 | 35,29 | 11,76 | 52,94 | 0 | 0 |
| Mail | 11,76 | 5,88 | 0 | 76,47 | 5,88 | 0 |
| Face-to-face | 0 | 21,43 | 35,71 | 42,86 | 0 | 0 |

**Table 3**: Communication tools (in % of observed ICLS).

By observing table 4 ,we can see that most of time, observed ICLS did not provide sharing/production tools even for tutor and learners (more than 80% of ICLS did not provided application sharing, whiteboard, shared text editor, wiki, blog and portfolio). Except for application sharing, sharing/production tools are not used when they are not provided by instructional designers. When wiki and blog are provided they are never used. Shared text editor is rather used by learners (25% of ICLS provided it and they are often used, 6,25% provide it and they are not used), and by tutors (12,5+6,25=18,75% provide it to tutors and they used it little or often). Whiteboard is rather (little or very often) used: 25% by learners and 14,44% by tutors. Application sharing is a little used when it is provided: 6,25% by learners and 12,5% by tutors.

Learners and tutors use rather consciously synchronised sharing/production tools (application, text and whiteboard) so we can suppose that it is because they are prescribed by instructional designer and so the use is made during specific activities. They do not use these tools when they are not prescribed, they do not need to use these tools alone, to organise themselves.

Collective work management tools observations are reporting in table 5. Theses tools are most of time not provided and not used by learners and tutors. In learners cases, proportions of ICLS providing management tools are 43,75% for document management, 30,76% for specific tools, 26,67% for scheduling tools, and 13,34% for awareness. When they are provided to learners, scheduling tools are very often used whereas other tools are half used and half not. In tutors cases, ICLS providing management tools are 33,33% for document management, 28,54% for specific tools, 25% for scheduling tools, and 15,38% for awareness. Most of tools are used except specific tools. Paradoxically specific tools are used (7,14%) even if they are not provided, same remark for awareness tools. We can suppose that initial uses of awareness and specific tools are turned away or adapted by tutors to serve specific needs not covered by the system (for instance to evaluate which student is present and active, which groups are collaborating and how they have collaborate).

ICLS provide other resources: 42% of them provide paper resources which are often used, 64% provide online resources very often used and 20% provide interactive resources which are very often used.

| Learners' uses | Provided and not used | Provided and a little used | Provided and often used | Not provided and not used | Not provided and a little used | Not provided and often used |
|---|---|---|---|---|---|---|
| Application sharing | 0 | 6,25 | 0 | 87,5 | 6,25 | 0 |
| Whiteboard | 0 | 12,5 | 12,5 | 75 | 0 | 0 |
| Shared text editor | 6,25 | 0 | 25 | 68,75 | 0 | 0 |
| Wiki | 12,5 | 0 | 0 | 87,5 | 0 | 0 |
| Blog | 6,25 | 0 | 0 | 93,75 | 0 | 0 |
| Portfolio | 0 | 0 | 0 | 100 | 0 | 0 |
| **Tutors' uses** | **Provided and not used** | **Provided and a little used** | **Provided and often used** | **Not provided and not used** | **Not provided and a little used** | **Not provided and often used** |
| Application sharing | 0 | 12,5 | 0 | 87,5 | 0 | 0 |
| Whiteboard | 0 | 6,67 | 6,67 | 86,66 | 0 | 0 |
| Shared text editor | 0 | 6,25 | 12,5 | 81,25 | 0 | 0 |
| Wiki | 50 | 0 | 0 | 50 | 0 | 0 |
| Blog | 6,25 | 0 | 0 | 93,75 | 0 | 0 |

**Table 4**: Sharing/production tools (in % of observed ICLS)

| Learners' uses | Provided and not used | Provided and a little used | Provided and often used | Not provided and not used | Not provided and a little used | Not provided and often used |
|---|---|---|---|---|---|---|
| Scheduling tool | 0 | 0 | 26,67 | 73,33 | 0 | 0 |
| Document management | 18,75 | 12,5 | 12,5 | 56,25 | 0 | 0 |
| Awareness | 6,67 | 0 | 6,67 | 86,66 | 0 | 0 |
| Specific tools | 15,38 | 0 | 15,38 | 69,24 | 0 | 0 |
| **Tutors' uses** | **Provided and not used** | **Provided and a little used** | **Provided and often used** | **Not provided and not used** | **Not provided and a little used** | **Not provided and often used** |
| Scheduling tool | 0 | 0 | 25 | 75 | 0 | 0 |
| Document management | 0 | 20 | 13,33 | 66,67 | 0 | 0 |
| Awareness | 0 | 7,69 | 7,69 | 76,92 | 7,69 | 0 |
| Specific tools | 7,14 | 7,14 | 7,14 | 71,43 | 7,14 | 0 |

**Table 5**: Collective work management tools (in % of observed ICLS)

## Conclusion

In this paper we have defined an ICLS (Instrumented Collective Learning Situations) as an educational situation with a learning goal (of knowledge or competencies), involving identified actors during a specified period of time and with an assessment of learners. We have observed, by the way of a survey, how ICLS are used in various e-learning contexts in France, Switzerland and Canada.

First results show that ICLS are not as common as initially supposed. Indeed, teaching educational models are based on learners' communication during individual tasks rather than on real collective tasks between learners. We can see that there is a gap between pedagogical researcher's recommendation (clearly mentioning collaboration need in e-learning activities) and practical e-learning set up.

When ICLS are used, a variety of collective activities are proposed to learners but cooperative vs. collaborative notions are not clearly differentiated. Activities are generally correctly achieved by learners and tutors. In terms of communication tools, (1) asynchronous ones are often used (forum, e-mail), (2) synchronous ones are, for one part of them (chat) often little used even if they are very often provided, and for another part (phone) often used (only between tutors). Concerning chat, it's important to note that chat tool itself is not prohibited because of the existence of external chat or instant messaging tools. It let us suppose that (1) learners want there communications to be private or (2) pedagogical activities using chat are not clearly expressed in scenarios. In more than 80% of observed ICLS sharing/production tools (especially blog, wiki and portfolio) are not provided for tutor and learners. When sharing/production tools are proposed (shared text editor, whiteboard or application sharing) that is to support a specific pedagogical activity; not to help actors to do there work. Collective work management tools are rather provided and used by learners and tutors and we have observed, in some cases, some uses appear whereas tools are not prescribed. We suppose that actors turn away or adapt theses tools for other uses (for instance to evaluate a learner's activity in a group).

In conclusion, we can say that collaboration is not really used and formalized in e-learning contexts. When there is collaboration, it is more communication between actors than a real collaborative work. Our future analyses will try to determine if instructional objectives are linked to specific tools and if they are really used. Our aim is to provide guidelines to instructional designers and tutors, helping them in their practices.

## References

Bannon, L.J. (1989). Issues in Computer-Supported Collaborative Learning, C. O'Malley, (Ed.), NATO *Advanced Workshop on Computer-Supported Collaborative Learning*, Maratea, Italy.

Bourriquen, B., David, J.-P., Garrot, E., George, S., Godinet, H., Medélez, E., & Metz, S. (2006). Caractérisation des Situations d'Apprentissage Collectives et Instrumentées dans le supérieur. *8ème Biennale de l'éducation et de la formation*, Lyon, France.

Casey, J., Brosnan, K., & Greller, W. (2005). Prospects for using learning objects and learning design as staff development tools in higher education. *IADIS International Conference on Cognition and Exploratory Learning in Digital Age (CELDA 2005)*, Porto, Portugal, 96-104.

Dillenbourg, P. (1999). What do you mean by 'collaborative learning'?, in P. Dillenbourg (Ed.). *Collaborative-learning: Cognitive and Computational Approaches*, Oxford, Elsevier, 1-19.

Doise W., & Mugny G. (1984). *The Social Development of the Intellect*. New York: Pergamon Press.

Garrot, E., George, S.,& Prévôt P. (2006) A System to Support Tutors in Adapting Distance Learning Situations to Students. *International Conference on Web Information Systems (WEBIST 2006)*, Setúbal, Portugal, 261-267.

George, S., & Leroux, P. (2001). Project-Based Learning as a Basis for a CSCL Environment: An Example in Educational Robotics. *In First European Conference on Computer-Supported Collaborative Learning (Euro-CSCL 2001), 2001*. Maastricht, The Netherlands, 269-276.

IMS (2003). *IMS Learning Design Information Model – version 1.0*. IMS Global Learning Consortium Inc., Retrieved October 24, 2006 from http://www.imsglobal.org/learningdesign/index.html

ISLS (2006). *International Society of the Learning Sciences. CSCL 2007 conference homepage*. Retrieved October 25, 2006 from http://www.isls.org/cscl2007/index.html

Koper R. (2000). *From change to renewal: Educational technology foundations of electronic learning environments*. Report from the Open University of the Netherlands, Retrieved October 24, 2006 from http://dspace.ou.nl/bitstream/1820/38/2/koper-inaugural-address-eng.pdf, 50 p.

McPherson, M., & Nunes, M. B. (2004). The role of tutors as an integral part of online learning support, *European Journal of Open, Distance and E-Learning (EURODL)*, 1, Retrieved October 24, 2006 from http://www.eurodl.org/materials/contrib/2004/Maggie_MsP.html

Jermann P., Soller A., & Muehlenbrock M. (2001). From mirroring to guiding: A review of state of the art technology for supporting collaborative learning. *European Conference on Computer-Supported Collaborative Learning* (Euro-CSCL 2001), Maastricht, Netherlands, 324-331.

Rabardel, P. (1995). *Les Hommes et les technologies. Approche cognitive des instruments contemporains,* Armand Colin.

## Acknowledgements

This work has been done in the context of ACTORS TICE Project financed by French Ministry of National Education and Research (action « Fields, technologies and theories », 2005-2007).